\newlength{\absize}

\documentclass[12pt]{article}
\usepackage{psfig}
\usepackage{epsf}

\renewcommand{\baselinestretch}{1.5}
\renewcommand{\arraystretch}{1.5}
\begin{document}
\thispagestyle{empty}
\pagestyle{empty}
\renewcommand{\thefootnote}{\fnsymbol{footnote}}
\newcommand{\starttext}{\newpage\normalsize
\pagestyle{plain}
\setlength{\baselineskip}{3ex}\par
\setcounter{footnote}{0}
\renewcommand{\thefootnote}{\arabic{footnote}}
}

\newcommand{\preprint}[1]{\begin{flushright}
\setlength{\baselineskip}{3ex}#1\end{flushright}}
\renewcommand{\title}[1]{\begin{center}\LARGE
#1\end{center}\par}
\renewcommand{\author}[1]{\vspace{2ex}{\Large\begin{center}
\setlength{\baselineskip}{3ex}#1\par\end{center}}}
\renewcommand{\thanks}[1]{\footnote{#1}}
\renewcommand{\abstract}[1]{\vspace{2ex}\normalsize\begin{center}
\centerline{\bf Abstract}\par\vspace{2ex}\parbox{\absize}{#1
\setlength{\baselineskip}{2.5ex}\par}
\end{center}}

\newcommand{\rep}{representation}
\newcommand{\tr}{\mathop{\rm tr}}
\newcommand{\cO}{{\cal O}}
\newcommand{\half}{{1\over2}}
\newcommand{\gtrsim}{\raisebox{.2em}{$\rlap{\raisebox{-.5em}{$\;\sim$}}>\,$}}
\newcommand{\ltsim}{\raisebox{.2em}{$\rlap{\raisebox{-.5em}{$\;\sim$}}<\,$}}
\newlength{\eqnparsize}
\setlength{\eqnparsize}{.95\textwidth}
\newcommand{\eqnbox}[1]{\parbox{\eqnparsize}{\bf\vskip.25ex #1\vskip1ex}}
\newcommand{\PSbox}[3]{\mbox{\rule{0in}{#3}\includegraphics{#1}\hspace{#2}}}

\setlength{\jot}{1.5ex}
\newcommand{\figsize}{\small}
\renewcommand{\bar}{\overline}
\font\fiverm=cmr5
\input prepictex
\input pictex
\input postpictex
\input{psfig.sty}
\newdimen\tdim
\tdim=\unitlength
\def\stpltsmbl{\setplotsymbol ({\small .})}
\def\bsmbl{\setplotsymbol ({\Huge .})}
\def\tarrow{\arrow <5\tdim> [.3,.6]}
\def\barrow{\arrow <8\tdim> [.3,.6]}

\setcounter{bottomnumber}{2}
\setcounter{topnumber}{3}
\setcounter{totalnumber}{4}
\renewcommand{\bottomfraction}{1}
\renewcommand{\topfraction}{1}
\renewcommand{\textfraction}{0}

\def\draft{\renewcommand{\label}[1]{{\quad[\sf ##1]}}
\renewcommand{\ref}[1]{{[\sf ##1]}}
\renewenvironment{equation}{$$}{$$}
\renewenvironment{thebibliography}{\section*{References}}{}
\renewcommand{\cite}[1]{{\sf[##1]}}
\renewcommand{\bibitem}[1]{\par\noindent{\sf[##1]}}}

\newcommand{\ximn}{\xi_{M/N}}
\def\cD{{\cal D}}
\def\cU{{\cal U}}

\def\D{ {\cal D}}
\def\U{ {\cal U}}
\def\Q{ {\cal Q}}
\newcommand\etal{{\it et al.}}

\newcommand{\outa}[3]{\startrotation by #1 #2 about 0 0
\tarrow from 15 0 to 32 0
\plot 32 0 45 0 /
\put {#3} at 60 0
\stoprotation}
\newcommand{\ina}[3]{\startrotation by #1 #2 about 0 0
\tarrow from 45 0 to 28 0
\plot 28 0 15 0 /
\put {#3} at 60 0
\stoprotation}
\newcommand{\outb}[3]{\startrotation by #1 #2 about 200 0
\tarrow from 215 0 to 232 0
\plot 232 0 245 0 /
\put {#3} at 260 0
\stoprotation}
\newcommand{\inb}[3]{\startrotation by #1 #2 about 200 0
\tarrow from 245 0 to 228 0
\plot 228 0 215 0 /
\put {#3} at 260 0
\stoprotation}

\def\theequation{\thesection.\arabic{equation}}
\preprint{\#HUTP-00/A004\\ 5/00}
\title{A Topcolor Jungle Gym\thanks{Research supported in
part by the
National Science Foundation
under grant number NSF-PHY/98-02709.}}
\author{
Howard Georgi\thanks{georgi@physics.harvard.edu} and Aaron K.
Grant\thanks{grant@carnot.harvard.edu} \\
Lyman Laboratory of Physics \\
Harvard University \\
Cambridge, MA 02138
}
\date{2/00}
\abstract{We discuss an alternative to the topcolor seesaw mechanism.
In our scheme, all the light quarks carry topcolor, and there are many
composite $SU(2)$ doublets. This makes it possible to get the observed $t$
quark mass and observed $SU(2)\times U(1)$ breaking in a way that is quite
different from the classic seesaw mechanism. We discuss a model of this
kind that arises naturally in the context of dynamically broken
topcolor. There are many composite scalars in a theory of this kind. This
has important effects on the Pagels-Stokar relation and the Higgs mass. We find $m_{\rm Higgs}\ltsim330$~GeV, lighter than in typical topcolor models. We
also show that the electroweak singlet quarks in such a model can be lighter
than the corresponding quarks in a seesaw model.  }
\starttext

\setcounter{section}{0}
\setcounter{equation}{0}

\section{Topcolor\label{introduction}}

We have known for many years that the weak interactions are associated with
the exchange of a massive gauge boson associated with a spontaneously broken
$SU(2)\times U(1)$ symmetry. The Goldstone bosons of the symmetry breaking are
eaten by the Higgs mechanism to become the longitudinal components of the
massive gauge bosons. But the nature of the Goldstone bosons remains
mysterious. Although there are fascinating hints that the Goldstone bosons
might be fundamental bosons from grand unified 
supermultiplets~\cite{Wilczek:1999tn}, it is
still possible that some more complicated dynamics is involved. The simplest
versions of strong dynamics are ruled out~\cite{technicolor}, so we know that
if a strong coupling scheme is to work, it must be very different from QCD.
 
Topcolor~\cite{topcolor} is a speculative scheme that may produce composite
Goldstone bosons and a composite Higgs boson~\cite{ch}. The idea is that
if a strong topcolor gauge symmetry is spontaneously broken, there is a
balance between the strong attractive gauge interaction that without
spontaneous breaking would confine topcolored particles, and the effect of
spontaneous breaking that liberates the topcolored particles by decreasing the
range of the interaction in the Higgs phase. The hope is that the transition
from the confining phase to the Higgs phase is smooth. If so, then we should
be able to tune the strength of the symmetry breaking to produce a light
composite scalar multiplet built out of the topcolored fermions. This
composite multiplet can then contain the
Goldstone bosons and a Higgs boson. There is no completely convincing proof of
the required smoothness, but it is consistent with everything we know.\footnote{Here we are assuming that the Coleman-Weinberg instability~\cite{cw} is under control --- we discuss this in section~\ref{higgsmass}. }

In section \ref{versus}, we introduce our new topcolor model (which we call
a ``topcolor jungle gym'') and compare it with a classic seesaw model. The
most obvious difference is that we have far more composite scalars below
the topcolor scale. In the remaining sections, we discuss a variety of
phenomenological issues that arise because of the large number composite
scalars in our model. We discuss flavor-changing neutral current effects in
section~\ref{fv}, the Pagels-Stokar relation in section~\ref{psr},
phenomenological constraints on the singlet quark masses in
section~\ref{singlets}, the size of the topcolor scale in
section~\ref{scales} and the Higgs mass and the stability of the vacuum in
section~\ref{higgsmass}. In appendix~\ref{flavor}, we work out a toy
example of some of the physics that could give rise to quark masses.

\section{Seesaw versus Jungle Gym\label{versus}}

In the classic seesaw model~\cite{seesaw}, the left-handed $(t,b)$ doublet and
a heavy right-handed particle, $\chi$, carry topcolor, and the Higgs multiplet
is a bound state of these two. The mass of the $t$ results from mixing between
the $t$ and the $\chi$, with the $t$ quark mass inversely related to the $\chi$
mass --- hence the name. This scheme has some advantages~\cite{dcgh}, but is
certainly not so compelling that we should ignore other possibilities.

In this note, we suggest a different scheme that emerged in our study of the
possibility that topcolor itself may be broken
dynamically~\cite{Collins:1999cf}. With
three or
more heavy particles like the $\chi$ of the conventional seesaw model we have
the option of having all the
observed quarks carrying topcolor, as shown in moose notation~\cite{moose} 
in figure~\ref{fig-3}. 

{\figsize\begin{figure}[htb]
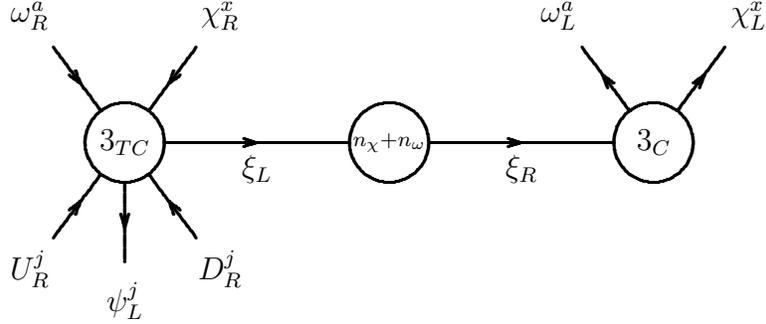

$$\beginpicture
\setcoordinatesystem units <\tdim,\tdim>
\stpltsmbl
\circulararc 360 degrees from 215 0 center at 200 0
\circulararc 360 degrees from 115 0 center at 100 0
\circulararc 360 degrees from 15 0 center at 0 0
\put {$3_{TC}$} at 0 0
\put {$3_C$} at 200 0
\put {\scriptsize $n_\chi$+$n_\omega$} at 100 0
\tarrow from 15 0 to 50 0
\plot 50 0 85 0 /
\tarrow from 115 0 to 150 0
\plot 150 0 185 0 /
\put {$\xi_R$} at 150 -10 
\put {$\xi_L$} at 50 -10 
\outa{0}{-1}{$\psi^j_L$}
\ina{.6}{.8}{$\chi^{x}_R$}
\ina{-.6}{.8}{$\omega^{a}_R$}
\outb{.6}{.8}{$\chi^{x}_L$}
\outb{-.6}{.8}{$\omega^{a}_L$}
\ina{-.6}{-.8}{$U^{j}_R$}
\ina{.6}{-.8}{$D^{j}_R$}
\linethickness=0pt
\putrule from 0 -75 to 0 75 
\endpicture$$
\caption{\figsize\sf\label{fig-3} A Jungle-Gym Moose.}\end{figure}}

In a conventional notation, the moose of figure~\ref{fig-3} describes the
quarks of the standard model, the electroweak doublets $\psi^j_L$, and the
singlets $U^j_R$ and $D^j_R$ for $j=1$ to 3, along with $n_\chi$ additional
charge $2/3$
quarks, $\chi^x$ for $x=1$ to $n_\chi$, and $n_\omega$ additional charge $-
1/3$ quarks, $\omega^a$ for $a=1$ to $n_\omega$. The gauge group is 
\begin{equation}
SU(3)_C\times SU(3)_{TC}\times SU(n_\chi\mathord{+}n_\omega)\times SU(2)\times U(1)
\label{gaugegroup}
\end{equation} 
The transformation properties of the fermion fields (the directed lines in the
moose diagram) are
\begin{equation}
\begin{array}{r@{\,:\;}l}
\psi^j_L&(1,3,1,2)_{1/6}\\
U^j_R&(1,3,1,1)_{2/3}\\
D^j_R&(1,3,1,1)_{-1/3}\\
\chi^x_R&(1,3,1,1)_{2/3}\\
\omega^a_R&(1,3,1,1)_{-1/3}\\
\end{array}
\quad\quad
\begin{array}{r@{\,:\;}l}
\xi_L&(3,1,n_\chi\mathord{+}n_\omega,1)_y\\
\xi_R&(1,3,n_\chi\mathord{+}n_\omega,1)_y\\
\chi^x_L&(3,1,1,1)_{2/3}\\
\omega^a_L&(3,1,1,1)_{-1/3}\\
\end{array}
\label{transform}
\end{equation}
where to cancel anomalies we need
\begin{equation}
y=-{2n_\chi-n_\omega\over3(n_\chi\mathord{+}n_\omega)}.
\end{equation}
This is a ``universal coloron'' model~\cite{uc} in the sense that all of
the light quarks carry the topcolor interactions. The $\chi$ and $\omega$
quarks are required for anomaly cancellation, and as we will see, play a
role in making the $t$ quark light. It is important that
$n_\chi$+$n_\omega$ be greater than or equal to three, and that $n_\chi$
and $n_\omega$ be greater than zero, but otherwise, these are not tightly
constrained.

In this model, the $\chi$ and $\omega$ play no direct role in the generation
of mass for the observed quarks. Instead, among the many composite Higgs
doublets are the following:
\begin{equation}
H_U^{m\bar{m}} \sim \bar{U}_R^{\bar{m}} \psi_L^m,
\quad\quad
H_D^{m\bar{m}} \sim \bar{D}_R^{\bar{m}} \psi_L^m.
\label{quarkhiggs}
\end{equation}
These have strong couplings to the quarks of which they are made with Yukawa
couplings of the form
\begin{equation}
h\left(\bar{\psi}^m_L\,
H_U^{m\bar{m}}
\,U^{\bar{m}}_R
+\bar{\psi}^m_L\,
H_D^{m\bar{m}}
\,D^{\bar{m}}_R\right)+\mbox{h.c.}
\label{strongyukawas}
\end{equation}
for large $h$ (in naive dimensional analysis, $h$ is of order $4\pi/\sqrt3$).

If the appropriate linear combination of these Higgs fields gets light and
develops a vacuum expectation value (VEV), the strong Yukawa couplings
produce the masses directly. The $\chi$ and $\omega$ may be important
because Higgs doublets built out of them can provide additional breaking of
the electroweak symmetry, to allow the Pagels-Stokar~\cite{ps} formula to
work for the $t$ quark (assuming, as we usually do for no very good reason,
that the naive Pagels-Stokar formula is accurate --- see
section~\ref{psr}).

The
mass matrix for the conventional seesaw model must get contributions from
at least three different kinds of operators and couplings.  However in the
model of figure~\ref{fig-3}, there are only two kinds of terms because
there is no difference except convention between $\chi_R$ and $U_R$. They
transform in the same way under the topcolor and electroweak gauge
symmetries, as you can see from (\ref{transform}). The mass matrix looks
generically like
\begin{equation}
\pmatrix{
\bar{\chi_R}&\bar{U_R}\cr
}
\,
\pmatrix{
x&y'\cr
x'&y\cr
}
\,
\pmatrix{
\chi_L\cr
U_L\cr
},
\label{m2}
\end{equation}
where the repeated letters are intended to indicate that the mechanisms are
the same for the $x$ and $x'$ and (separately) for the $y$ and $y'$.  By
purely conventional redefinitions of what is $\chi_R$ and and what is $U_R$
we can put zeros in (\ref{m2}) and write it either as\footnote{When we make
these redefinitions, the numerical values of the matrices $x$, $y$, $x'$
and $y'$ will change.}
\begin{equation}
\pmatrix{
\bar{\chi_R}&\bar{U_R}\cr
}
\,
\pmatrix{
x&y'\cr
0&y\cr
}
\,
\pmatrix{
\chi_L\cr
U_L\cr
},
\label{m3}
\end{equation}
or as
\begin{equation}
\pmatrix{
\bar{\chi_R}&\bar{U_R}\cr
}
\,
\pmatrix{
x&0\cr
x'&y\cr
}
\,
\pmatrix{
\chi_L\cr
U_L\cr
}.
\label{m4}
\end{equation}
The form in (\ref{m3}) is often the most useful, because we are typically
interested in $x\gg y$, so it pays to diagonalize $x$.

There are a number of important consequences of the form (\ref{m3}). 

First, we notice that all of the quarks can get masses from the dimension 6
operators. Of course, this is both good news and bad news. It means that we
have a potentially realistic model, but also that there is no trivial
explanation of the hierarchy of quark masses. This must come from the
details of the flavor physics.

Second, the decoupling limit of the theory is very straightforward. If $x\gg
y$, the light quark mass matrix is determined simply by the lower right-hand
corner: 
\begin{equation}
\pmatrix{
\bar{\chi_R}&\bar{U_R}\cr
}
\,
\pmatrix{
x&y'\cr
0&\fbox{$y$}\cr
}
\,
\pmatrix{
\chi_L\cr
U_L\cr
}.
\label{m5}
\end{equation}
The $y'$ matrix in the upper right hand corner plays no role in the fermion mass spectrum in this
limit, however, it is important in the Pagels-Stokar relation because the
composite doublets whose VEVs contribute to this term also contribute to
electroweak symmetry breaking.

Of course, the decoupling limit is also possible for the mass matrix of the
conventional seesaw model, but it requires some tuning of the various
different types of contributions to maintain the relation between the
different types of terms in the mass matrix.

Note that because the difference between the $U_R$ and $\chi_R$ fields, and
between the $D_R$ and $\omega_R$ fields is purely conventional (depending
on the details of the flavor physics), it can be useful psychologically to
combine them into multiplets, $\cU_R$ and $\cD_R$, and display the Moose as
in figure~\ref{fig-4}.  
{\figsize\begin{figure}[htb]
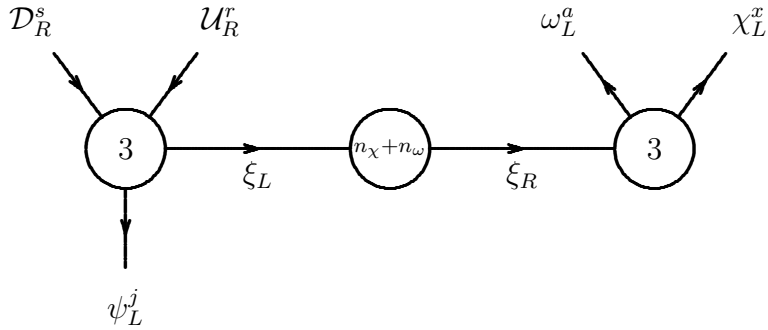

$$\beginpicture
\setcoordinatesystem units <\tdim,\tdim>
\stpltsmbl
\circulararc 360 degrees from 215 0 center at 200 0
\circulararc 360 degrees from 115 0 center at 100 0
\circulararc 360 degrees from 15 0 center at 0 0
\put {$3$} at 0 0
\put {$3$} at 200 0
\put {\scriptsize $n_\chi$+$n_\omega$} at 100 0
\tarrow from 15 0 to 50 0
\plot 50 0 85 0 /
\tarrow from 115 0 to 150 0
\plot 150 0 185 0 /
\put {$\xi_R$} at 150 -10 
\put {$\xi_L$} at 50 -10 
\outa{0}{-1}{$\psi^j_L$}
\ina{.6}{.8}{$\cU^r_R$}
\ina{-.6}{.8}{$\cD^s_R$}
\outb{.6}{.8}{$\chi^{x}_L$}
\outb{-.6}{.8}{$\omega^{a}_L$}
\linethickness=0pt
\putrule from 0 -75 to 0 75 
\endpicture$$
\caption{\figsize\sf\label{fig-4} The Moose of 
figure~\protect\ref{fig-3} in a notation in which the $U_R^j$ and
$\chi_R^x$ fields are combined into a $\cU_R^r$ multiplet (where $r$ runs
from 1 to $n_\chi$+3), and the $D_R^j$ and $\omega_R^a$ fields are combined
into a $\cD_R^s$ multiplet (where $s$ runs from 1 to
$n_\omega$+3).}\end{figure}}

The flavor physics in the model of figure~\ref{fig-3} (or equivalently,
figure~\ref{fig-4}) can be rather simple. No mass terms are allowed. One needs
five types of 4-fermion operators ---
\begin{eqnarray}
&\displaystyle 
\bar{\chi^x_L}\,\xi_R\;\bar{\xi_L}\,\cU^r_R,
&\label{fn1}\\
&\displaystyle 
\bar{\omega^x_L}\,\xi_R\;\bar{\xi_L}\,\cD^s_R,
&\label{fn2}\\
&\displaystyle 
\bar{\psi^j_L}\,\cU^r_R\;\bar{\psi^{j'}_L}\,\cD^s_R,
&\label{fn3}\\
&\displaystyle 
\bar{\psi^j_L}\,\cU^r_R\;\bar{\cU^{r'}_R}\,\psi^{j'}_L,
&\label{fn4}\\
&\displaystyle 
\bar{\psi^j_L}\,\cD^s_R\;\bar{\cD^{s'}_R}\,\psi^{j'}_L.
&\label{fn5}
\end{eqnarray}
The terms in (\ref{fn4}) and (\ref{fn5}) can arise from flavor gauge
interactions, but (\ref{fn1}-\ref{fn3}) must come from more complicated
flavor physics. Of the latter, (\ref{fn1}) and (\ref{fn2}) have a very
simple structure and can be simply parameterized. On the other hand, the
Peccei--Quinn-symmetry breaking terms in (\ref{fn3}) involve a very large
number of parameters. It may be possible to assume that all these terms are
small. In appendix~\ref{flavor}, we work out a specific example in which
the terms in (\ref{fn4}) and (\ref{fn5}) come from an $SO(3)$ gauge
interaction.  While we are unable to obtain a realistic mass matrix in this
simple scheme, we do find a rich structure that we find encouraging.

\section{Phenomenology\label{phenomenology}}

\subsection{Flavor Violation\label{fv}}

Below the topcolor scale, the $SU(3)\times SU(2)\times U(1)$ structure of
this model is similar to that of a standard seesaw model. The differences
here are that there are additional singlet quarks and more composite
scalars. One might worry that the additional strongly coupled 
composite scalars will produce
a host of unwanted flavor violating effects. However, this need not be the
case. These effects will depend on the details of the flavor physics. The
reason is that in the absence of flavor physics, the model has a large
non-Abelian flavor symmetry acting on the topcolored quarks,
\begin{equation}
SU(3)_L\times SU(3\mathord{+}n_\chi)_U\times SU(3\mathord{+}n_\omega)_D,
\label{fsyms}
\end{equation}
with the factors acting on the $\psi_L$, the $\cU_R=U_R\mathord{+}\chi_R$
and the $\cD_R=D_R\mathord{+}\omega_R$ multiplets, respectively.  In fact,
the symmetry of the strong interactions is still larger,
\begin{equation}
SU(6)_L\times SU(6\mathord{+}n_\chi\mathord{+}n_\omega)_{U+D},
\label{fsyml}
\end{equation}
but (\ref{fsyml}) is broken down to (\ref{fsyms}) by the $SU(2)\times U(1)$
electroweak gauge interactions.  The composite scalar doublets carry the
transformation properties of the quarks out of which they are made,
thus\footnote{Note that (\ref{hu}) and (\ref{hd}) are the analog of
(\ref{quarkhiggs}) with $U$ and $D$ replaced by $\U$ and $\D$.}
\begin{equation}
H_\U^{m\bar{m}} \sim \bar{\U}_R^{\bar{m}} \psi_L^m
\label{hu}
\end{equation}
transforms like an electroweak doublet (with hypercharge $Y=-1/2$) and like
$(3,\bar{3\mathord{+}n_\chi})$ under (\ref{fsyms})
and
\begin{equation}
H_\D^{m\bar{m}} \sim \bar{\D}_R^{\bar{m}} \psi_L^m
\label{hd}
\end{equation}
transforms like an electroweak doublet (with $Y=1/2$) and like
$(3,\bar{3\mathord{+}n_\omega})$ under (\ref{fsyms}). The combination,
\begin{equation}
H=\pmatrix{
H_\U&H_\D\cr
}
\label{hud}
\end{equation}
transforms like a $(6,\bar{6\mathord{+}n_\chi\mathord{+}n_\omega})$ under
(\ref{fsyml}). In terms of (\ref{hud}), the Yukawa couplings to the quarks
look like
\begin{equation}
h\,\bar\psi_L\,H\,\pmatrix{
\U_R\cr
\D_R\cr
}+\mbox{h.c.}
\label{yukawa}
\end{equation}

In particular the $H_\U$ and $H_\D$ composite scalar doublets are degenerate
to lowest order in the absence of flavor physics. Now suppose that the
flavor physics picks out a single linear combination of these scalar
doublets and makes it light - with a slightly negative mass squared, so
that this linear combination contains the Higgs boson and the Goldstone
bosons of the standard model. Now we will show that all the extra doublets
really do not introduce any additional flavor violation beyond that
associated in a well defined sense with the exchange of the single Higgs
multiplet. This is an overly simplistic picture of what the flavor physics
does, to be sure, but it will illustrate the point that the extra flavor
symmetry will in general suppress flavor changing effects. The idea of the
demonstration is simple. Because of the large flavor symmetry, composite
scalar exchange does not produce flavor violation if the scalars are all
degenerate. If one linear combination has a different mass, it is only that
difference that produces flavor violation. We will show more formally how
this works below.

In general, we can write the properly normalized linear combination of scalar
doublets that gets light as
\begin{equation}
\phi_1=\cos\theta\,H_u+\sin\theta\,\tilde H_d,
\label{phi1}
\end{equation} 
where
\begin{equation}
\tilde\phi\equiv i\tau_2\,\phi^*,
\end{equation}
and
\begin{equation}
H_u\equiv\tr\Bigl(u^\dagger\,H_\U\Bigr)
\quad\mbox{and}\quad
H_d\equiv\tr\Bigl(d^\dagger\,H_\D\Bigr),
\label{huudd}
\end{equation}
for a $3$$\times$$(3\mathord{+}n_\chi)$ matrix $u$ and a 
$3$$\times$$(3\mathord{+}n_\omega)$ matrix $d$ satisfying
\begin{equation}
\tr\Bigl(u\,u^\dagger\Bigr)
=\tr\Bigl(d\,d^\dagger\Bigr)=1.
\label{udnorm}
\end{equation}
It is useful to define also the following fields:
\begin{equation}
\phi_2=\sin\theta\,H_u-\cos\theta\,\tilde H_d,
\label{phi2}
\end{equation} 
\begin{equation}
H_{\hat{\U}}\equiv H_\U-u\,H_u,
\quad\quad
H_{\hat{\D}}\equiv H_\D-d\,H_d.
\label{hhatuandd}
\end{equation}
In terms of these fields, we can write
\begin{equation}
H_\U=H_{\hat{\U}}+u\,(\cos\theta\,\phi_1+\sin\theta\,\phi_2),
\label{huphi}
\end{equation}
\begin{equation}
H_\D=H_{\hat{\D}}+d\,(\cos\theta\,\tilde\phi_2-\sin\theta\,\tilde\phi_1).
\label{hdphi}
\end{equation}
The important point about (\ref{huphi}) and (\ref{hdphi}) is they show exactly
how $\phi_1$ appears in the matrix structure of (\ref{yukawa}).

Now to see the result, it is convenient to add two more fields, both with
exactly the same couplings as $\phi_1$, call them $\phi'_1$ and $\phi''_1$.
Both $\phi'_1$ and $\phi''_1$ are degenerate with $\phi_2$ and all the
other composite doublets except $\phi_1$. But we choose $\phi''_1$ to be a
ghost field, so that its effect cancels that of $\phi'_1$. The point is
that we can now group $\phi'_1$ with $\phi_2$, $H_{\hat{\U}}$ and
$H_{\hat{\D}}$ into a complete degenerate multiplet under the (\ref{fsyml})
flavor symmetry. Thus the couplings of these scalars produce no flavor
change. All the flavor change comes from the propagation of $\phi_1$
and $\phi''_1$, which have the same couplings. This is the sense in which,
as promised, the flavor violation is related to that produced by $\phi_1$
alone.

\subsection{The Pagels-Stokar relation\label{psr}}
In this section we discuss the constraint on the quark mass matrix in this
model coming from the analog of the Pagels-Stokar relation. In this
context, the Pagels-Stokar relation is essentially a formula for the Yukawa
couplings of the composite Higgs. In this model, because of the large
flavor symmetry, (\ref{fsyml}), it becomes a relation for the sum of the
squares of all the terms in the quark mass matrix involving the left handed
quark doublets. We can see this very simply using the formalism developed
in section \ref{fv}.

The masses of the charge $2/3$ and charge $-1/3$ quarks have the following
form:
\begin{equation}
\pmatrix{
\bar{\cU_R}\cr
}
\,
\pmatrix{
X_u&Y_u\cr
}
\,
\pmatrix{
\chi_L\cr
U_L\cr
},
\quad\quad
\pmatrix{
\bar{\cD_R}\cr
}
\,
\pmatrix{
X_d&Y_d \cr
}
\,
\pmatrix{
\omega_L\cr
D_L\cr
},
\label{psmm}
\end{equation}
where $X_u$ (a $(3$+$n_\chi)$$\times$$n_\chi$ matrix) and $X_d$ (a
$(3$+$n_\omega)$$\times$$n_\omega$ matrix) are the contributions to the
singlet masses and $Y_u$ (a $(3$+$n_\chi)$$\times$$3$ matrix) and $Y_d$ (a
$(3$+$n_\omega)$$\times$$3$ matrix) are contributions to the masses from
$SU(2)\times U(1)$ breaking.

Let $\phi_1$ from (\ref{phi1}) be the linear combination of composite Higgs
doublets that develops a vacuum expectation value.\footnote{Note that in
this section, unlike \ref{fv}, we are not assuming that only this linear
combination is light --- the discussion here only assumes that the vacuum
expectation value preserves the electromagnetic $U(1)$.} It is then clear
from (\ref{huphi}), (\ref{hdphi}) and (\ref{yukawa}) that the symmetry
breaking mass matrices have the form\footnote{We adopt a normalization
where the $SU(2)$ breaking VEV is $v=246$ GeV.}
\begin{equation}
Y_u=h\,u\,v\cos\theta/\sqrt2,
\quad\quad
Y_d=h\,d\,v\sin\theta/\sqrt2.
\label{yuandyd}
\end{equation}
Thus using (\ref{udnorm}) we immediately conclude 
\begin{equation}
\tr\Bigl(Y_u\,Y_u^\dagger\Bigr)
+\tr\Bigl(Y_d\,Y_d^\dagger\Bigr)=h^2\,v^2/2.
\label{h2}
\end{equation}

Thus the sum of the squares of the $SU(2)\times U(1)$ breaking mass terms
are determined by the square of coupling $h$. This coupling should be
evaluated at a scale of the order of $\Lambda\approx1$~TeV, the electroweak
breaking scale. If the coupling at the topcolor scale, $\Lambda_{TC}$, is
$h(\Lambda_{TC})$, then the coupling runs via the renormalization group
down to a smaller value at the scale $\Lambda$. In our theory, all of the
composite scalars in the low energy theory and all the quarks to which they
couple contribute to the running of $h$. The running is thus not the same
as in the simple seesaw model, and it depends on the masses of the
composite scalars and the singlet quarks. For simplicity, let us assume, as
we did in section \ref{psr}, that the composite scalars are approximately
degenerate except for the linear combination that becomes the Higgs
doublet. Call the common mass of the other scalars $m_s$. We will also
assume that the singlet quark mass terms are approximately equal. Call the
mass of the singlet quarks $m_{sf}$. We will see in section~\ref{scales}
that $m_{sf}$ is naturally smaller than $m_s$. Finally, we will assume (for
simplicity) that $m_{sf}$ is of order 1~TeV,
\begin{equation}
m_{sf}\approx\Lambda.
\label{allone}
\end{equation}
In section \ref{scales} we will estimate all these masses using dimensional
analysis, and we will see that this is the most interesting case.

Between the scale $\Lambda_{TC}$ where the composite scalars are bound and
the scale $m_s$, the theory has the full chiral flavor symmetry of
\begin{equation}
SU(6)_L\times SU(6\mathord{+}n_\chi\mathord{+}n_\omega)_{U+D}.
\label{fsyml2}
\end{equation}
Thus each complex component of the $6$ by
$6\mathord{+}n_\chi\mathord{+}n_\omega$ matrix composite scalar satisfies a
renormalization group equation with
\begin{equation}
\beta_h={h^3\over32\pi^2}(2N_C+N_L+N_R)
\label{beta}
\end{equation}
where 
\begin{equation}
N_L=6 \,,\quad\quad
N_R=6\mathord{+}n_\chi\mathord{+}n_\omega \,.
\label{beta2}
\end{equation}

Between the scale $m_s$ and the scale $m_{sf}$, what happens really depends
on the structure of the breaking of the flavor symmetry. It is conservative
(in the sense that it leads to slower running of the Yukawa coupling) to
assume that only the Higgs field survives in this region, in which case
$N_L=2$ and $N_R=1$. Because we have assumed that $m_{sf}$ is not so
different from $\Lambda\approx1$~TeV, we can use $h(m_{sf})$ in our
Pagels-Stokar relation.  Then we can write the renormalization group result
as
\begin{equation}
{32\pi^2\over h(m_{sf})^2}=
{32\pi^2\over h(\Lambda_{TC})^2}
+(2N_C+N_L+N_R)\ln(\Lambda_{TC}^2/m_s^2)
+(2N_C+3)\ln(m_s^2/m_{sf}^2).
\label{rg}
\end{equation}
Putting in $N_C=3$, (\ref{msf}), (\ref{mscalar}), (\ref{beta2}) and using
the mininum possible value of $n_\chi\mathord{+}n_\omega=3$, (\ref{rg})
becomes
\begin{equation}
{32\pi^2\over h(m_{sf})^2}=
{32\pi^2\over h(\Lambda_{TC})^2}
+21\ln(\Lambda_{TC}^2/m_s^2)
+9\ln(m_s^2/m_{sf}^2).
\label{rg2}
\end{equation}
To say more, we must know something more about the various scales in
(\ref{rg2}). We will discuss this in detail in section~\ref{scales}.

In the limit that $h(\Lambda_{TC})$ is very large, (\ref{rg2}) becomes
\begin{equation}
h(\Lambda)^2={32\pi^2\over21\ln(\Lambda_{TC}^2/m_s^2)
+9\ln(m_s^2/m_{sf}^2)}.
\label{npsr}
\end{equation}
The condition
\begin{equation}
h(\Lambda_{TC})=\infty
\label{compositeness}
\end{equation}
is sometimes called the compositeness condition because it can be naively
interpreted in terms of the vanishing of the wave function of the
fundamental Higgs scalar at the topcolor scale. In our view, this is not a
particularly useful way of thinking. Rather, we would argue that the finite
parameter $h(\Lambda_{TC})$ is an important, non-perturbative constant that
we need to learn how to calculate to make sense of the topcolor
theory. Until we can do that, we will adopt $h(\Lambda_{TC})=\infty$ and
(\ref{npsr}) as a provisional estimate, but we will also consider what
happens if $h(\Lambda_{TC})$ is smaller. Thus the analog of the
Pagels-Stokar relation is
\begin{equation}
\tr\Bigl(Y_u\,Y_u^\dagger\Bigr)
+\tr\Bigl(Y_d\,Y_d^\dagger\Bigr)={16\pi^2\,v^2\over21\ln(\Lambda_{TC}^2/m_s^2)
+9\ln(m_s^2/m_{sf}^2)},
\label{psp}
\end{equation}
but it should be interpreted as an upper bound on the $SU(2)\times U(1)$
symmetry breaking terms in the quark mass matrix.

Note that the right hand side of (\ref{psp}) is not the standard
Pagels-Stokar relation, which is calculated in the large $N_C$ limit -
ignoring $N_L$ and $N_R$, and thus has the form
\begin{equation}
h(\Lambda)^2={16\pi^2\,v^2\over3\ln(\Lambda_{TC}^2/\Lambda^2)}.
\label{opsr}
\end{equation}
In our model, the flavors make a contribution that should not be ignored.
The additional running caused by the extra composite scalar fields reduces 
the contribution to the quark masses. 

\subsection{Constraints on Singlets\label{singlets}}

In this section we discuss constraints on the masses and mixings of the
$SU(2)$ singlet quarks.  In the presence of these singlets, the mass eigenstate
quarks are in general a mixture of $SU(2)$ doublets and $SU(2)$ singlets,
leading to a variety of non-standard effects in the low-energy theory.  The
effect of the mixing is to modify the couplings of left-handed quarks to
the $SU(2)$ gauge bosons.  Since the extra right-handed quarks have the
same couplings as the standard ones, the mixing in this sector is
irrelevant.  If a standard quark $q$ mixes with a singlet with mixing angle
$\phi_q$, the left-handed quark's coupling to the $Z$ is
\begin{equation}
g_L = \cos^2 \phi_q ( I_3 - Q \sin^2 \theta_W ) 
+ \sin^2 \phi_q ( -Q\sin^2\theta_W ).
\end{equation}
Couplings of doublet quarks to the $W$ are suppressed by a factor $\cos^2
\phi_q$ by this mixing.  At tree level, mixing modifies $Z$ decay
widths and neutral current phenomena such as deeply inelastic
neutrino-nucleon scattering and parity violation in atoms.  Loop effects
can also be important.  In particular, if the mixing does not preserve a
custodial $SU(2)$ symmetry, the singlets can give large contributions to
the $\rho$ parameter \cite{seesaw,Collins:1999cf}.  In the charged-current
sector, the mixing can give rise to a non-unitary CKM matrix and anomalous
$W$ decay widths.

Constraints on mixing angles have been considered in detail in
ref.~\cite{Popovic:2000dx}.  The jungle gym model most nearly resembles
Model A of ref.~\cite{Popovic:2000dx}, where all the standard model
fermions have singlet partners.  Model A differs from the jungle gym since
both leptons and quarks have singlet partners.  At 95\% confidence level,
the constraints on mixing of the light quarks are \cite{Popovic:2000dx}
\begin{equation}
\label{MixingBounds}
\begin{array}{cc}
\sin^2\phi_u < 0.013, \ \sin^2\phi_d < 0.015, \\
\sin^2\phi_c < 0.020, \ \sin^2\phi_s < 0.015.
\end{array}
\end{equation}
Measurements of $R_b$ lead to a more stringent constraint on mixing of the
$b$ quark:
\begin{equation}
\sin^2\phi_b < 0.0025.
\end{equation}

Mixing of the top quark is constrained by loop effects.  The electroweak
$\rho$ parameter provides the most stringent bounds.  In the case of a
single doublet mixing with a heavy singlet, the contribution to $\rho$ is
\cite{seesaw}
\begin{equation}
\delta\rho = {N_c \over 16 \pi^2 v^2} 
\left[ \sin^4\phi_q m^2_\chi + 
2\sin^2\phi_q
\cos^2\phi_q {m^2_\chi m^2_q\over m^2_\chi - m^2_q}
\ln {m^2_\chi\over m^2_q}
-\sin^2\phi_q (2-\sin^2\phi_q) m_q^2 \right].
\end{equation}
This contribution to $\rho$ is positive and implies a bound on the mass and
mixing of the heavy singlet; in the simple seesaw case, the lower bound on
the $\chi$ mass is about 5~TeV~\cite{Collins:1999cf} if we use the naive
Pagels-Stokar relation.  The jungle gym case is more complicated.  If, for
instance, a pair of $\chi$ and $\omega$ quarks are degenerate and have equal
mixing with an $SU(2)$ doublet of light quarks, then the model preserves a
custodial $SU(2)$ symmetry, and the contribution to $\rho$ is greatly
suppressed \cite{Collins:1999cf}.

The exact bounds on the masses and mixings of the singlets depend upon a
number of factors.  Lower bounds on masses contain a factor of the topcolor
Yukawa coupling $h$.\footnote{There is also implicit dependence on $h$ that
we will discuss below.} Also, from the preceding paragraphs we see that the
mixing bounds vary significantly depending on the flavor of the quark that
condenses.  Typically, the bound on a singlet $Q$ will be given by
\begin{equation}
m_Q \geq \frac{ h \,v_{Qq}}{\sin\phi_q},
\label{sb}
\end{equation}
where $v_{Qq}$ is the expectation value of the composite field made of
$\bar Qq$,
\begin{equation}
v_{Qq}\equiv\langle H_{\bar{Q}q}\rangle.
\label{vsq}
\end{equation}
We can have lighter singlets by making the VEV $v_{Qq}$ smaller or by
choosing the flavor of $q$ to maximize $\sin\phi_q$.  One possibility would
be to distribute the $SU(2)$ breaking VEVs uniformly over all of the
composite Higgs fields $H_{u,d}$.  However, this alternative is likely to
lead to tree-level flavor changing neutral currents in $Z$ exchange, which
is clearly not permissible.  A second possibility is to give equal VEVs to
a pair of up- and down-type Higgses.  Depending on details of the flavor
physics, this need not lead to tree-level flavor changing neutral currents.
This pattern of VEVs has the added virtue that it results in a negligible
contribution to the $\rho$ parameter if the corresponding $\chi$ and
$\omega$ quarks are approximately degenerate.  The lightest possible
singlets result if we let the $(s,c)$ doublet of quarks condense.  One
might be tempted to have condensates of $\bar{t}_L \chi_R$ and $\bar{s}_L
\omega_R$, for instance, but in this case contributions to the $\rho$
parameter are not small: the condensing quarks must belong to the same
$SU(2)$ doublet to have a small value of $\delta\rho$.  Thus the least
constrained case seems to involve equal mixing of two degenerate singlets
with the $c$ and $s$ respectively.

Now let us discuss the implicit dependence of $v_{Qq}$ on $h$. We can
safely ignore all the quark masses in (\ref{h2}) except for
$m_t\approx174$~GeV. If we also assume that the only doublet-singlet mixing
arises from equal expectation values for mixing with $c$ and $s$, we have
\begin{equation}
v_{Qq}^2\approx\half(v^2/2-m_t^2/h(m_{sf})^2),
\label{cs}
\end{equation}
or 
\begin{equation}
m_{sf}^2\gtrsim \half \frac{(h(m_{sf})^2v^2/2-m_t^2)}{\sin^2 \phi_q},
\label{cs2}
\end{equation}
where $h(m_{sf})$ is given by (\ref{rg2}). This produces the implicit
dependence on $h(\Lambda)$ referred to above.

We will see in the next section that the jungle gym model admits lighter
singlets than the standard seesaw model, where one finds a bound in the
5-12 TeV range \cite{Collins:1999cf}.\footnote{If we allow for
$h(\Lambda)<\infty$, the bounds are relaxed in the seesaw model as well,
but are still higher than in the model we consider.}

The pattern of VEVs here is sufficiently bizarre that it merits further
discussion.  One might worry in particular that the mixing would result in
a violation of CKM unitarity that conflicts with experiment.  This is not
the case.  We can see this by considering mixing of the first two
generations.  It is safe to ignore the third generation since the mixing
with the third generation is small and since the mixing angles in this
sector are not as well measured as for the first two generations.  We can
derive the effect of the mixing with singlets by considering a two-step
rotation from the weak basis to the mass basis.  First, in the weak basis
the $W$ couples only to doublet quarks, and there are no
generation-changing couplings.  Now, perform $SU(3)_{L,R}$ rotations on the
standard model quarks to diagonalize their mass terms.  In this basis there
are still off-diagonal mass terms that mix doublets with singlets, and the
$W$ couples to the standard model quarks with a unitary CKM matrix.  By
assumption, only the $c$ and $s$ quarks mix with the singlets, so we can
now go to the final mass basis by performing rotations on $(c,\chi)_{L,R}$
and $(s,\omega)_{L,R}$.  The $W$ now couples to the standard quarks with a
non-unitary CKM matrix, and the mass terms are fully diagonalized.  The
mixing in the left-handed sector is set by the singlet mass and the $SU(2)$
breaking mass, which in turn is set by the Pagels-Stokar relation: the
left-handed $c$-mass eigenstate is given by
\begin{equation}
c_L^{\rm mass} = \cos\theta_L c_L + \sin\theta_L \chi_L,
\end{equation}
with 
\begin{equation}
\sin\theta_L \simeq \frac{ h \langle H_{\bar{\chi}c} \rangle }{ M_\chi }
\end{equation}
for $M_\chi \gg h \langle H_{\bar{\chi}c} \rangle$.  In terms of the
Cabbibo angle $\theta_C$ and the mixing angle $\theta_L$, the mixing matrix
for the two-generation case has the form
\begin{equation}
\label{TheoryCKM}
\left( \begin{array}{cc}
\cos\theta_C & \sin\theta_C \cos\theta_L \\
-\sin\theta_C \cos\theta_L & \cos\theta_C \cos^2\theta_L
\end{array} \right).
\end{equation}
This is to be compared with the {\it direct} determination of the CKM
matrix, where no assumption of unitarity has been made.  The Particle Data
Group \cite{PDG} gives
\begin{equation}
\label{ExperimentCKM}
\left( \begin{array}{cc}
0.9735 \pm 0.0008 & 0.2196 \pm 0.0023 \\
0.224  \pm 0.016  & 1.04   \pm 0.16 
\end{array} \right)
\end{equation}
for the magnitudes of the mixing angles.  Curiously, the best fit of
(\ref{TheoryCKM}) to (\ref{ExperimentCKM}) gives a large value of
$\theta_L$:  the best fit is obtained with
\begin{equation}
\theta_C = 0.2302,~~~\theta_L = 0.2700.
\end{equation}
The large value of $\theta_L$ means that there is no meaningful constraint
on the mixing from CKM unitarity.  The large value of $\theta_L$ comes
about because the first row of the CKM matrix (\ref{ExperimentCKM}) is not
particularly unitary:  we would expect that $|V_{ud}|^2 + |V_{us}|^2=1$, up
to very small corrections from $V_{ub}$.   Instead we find 
$|V_{ud}|^2 + |V_{us}|^2=0.9959\pm 0.0018$, roughly $2\sigma$ away from
unity.  As a result, $V_{ud}$ favors a larger value of $\theta_C$ than
$V_{us}$;  we can eliminate this disagreement with a non-zero value of
$\theta_L$.  The large error on $V_{cs}$ also permits a large value of
$\theta_L$.

We also note that the small ($\sim \theta_C$) mismatch between the weak-
and mass-eigenstate $c$ and $s$ quarks results in a slightly larger value
of $\delta\rho$ than one would find in the case of no mixing.  However, the
contribution to $\delta\rho$ from mixing is suppressed by $\sin^2\theta_C$
and so is small enough that the estimates we will discuss in the next
section are not noticeably affected.

\section{The Topcolor Scale\label{scales}}

So far, we have discussed only the low energy theory below the topcolor
scale. We do not know how to construct any complete model of the physics at
higher energies. However, we can use the results of the previous sections
and the tenets of effective field theory to develop a detailed picture of
the different scales involved. We will find that if we assume that all the
new physics beyond topcolor comes from a single large scale, $f_F$, then all
the scales in the theory will be approximately determined.

First let us make some general remarks. There are two kinds of fine tunings
required in a theory of this kind. For the model to be at all attractive,
both must be modest. One fine-tuning is the tuning of the common mass of
the composite scalars to be lower than the topcolor scale. This tuning is
the crux of the topcolor model, and involves (in our model with dynamically
broken topcolor) the relative strength of the two strong gauge groups at
the topcolor scale. The other tuning is required to make the electroweak
symmetry breaking vacuum expectation value small compared to the mass of
the composite scalars. This involves a tuning of the coefficient of the
dimension 6 operator responsible for splitting the Higgs multiplet from the
other composite scalars, giving it a negative mass squared smaller in
magnitude than the common mass from the topcolor physics. We will see that
neither of these tunings needs to be very fine. Furthermore, it will turn
out that it is very natural for both tunings to be at about the same
level. We are not sure how uncomfortable we should be that we need two of
them, or how to compare the ``naturalness'' of this scheme with models in
which there is a single much finer tuning. We will simply describe how it
works.

We listed in (\ref{fn1}-\ref{fn5}) the dimension six operators that appear
as the most important interactions in the effective theory below
$f_F$ from physics at higher scales. The singlet fermion masses
arise from (\ref{fn1}-\ref{fn2}). Mass splittings among the various
composite scalars arise directly from(\ref{fn3}-\ref{fn5}) and indirectly
from the singlet masses.

Let us assume that all of these operators appear in the effective theory as
nonrenormalizable interactions, suppressed by appropriate powers of some
large scale, $f_F$. The interaction terms are
\begin{eqnarray}
&\displaystyle 
{1\over f_F^2}\,\bar{\chi^x_L}\,\xi_R\;\bar{\xi_L}\,\cU^r_R,
&\label{ffn1}\\
&\displaystyle 
{1\over f_F^2}\,\bar{\omega^x_L}\,\xi_R\;\bar{\xi_L}\,\cD^s_R,
&\label{ffn2}
\end{eqnarray}
which contribute to the masses of the singlet fermions and
\begin{eqnarray}
&\displaystyle 
{1\over f_F^2}\,\bar{\psi^j_L}\,\cU^r_R\;\bar{\psi^{j'}_L}\,\cD^s_R,
&\label{ffn3}\\
&\displaystyle 
{1\over f_F^2}\,\bar{\psi^j_L}\,\cU^r_R\;\bar{\cU^{r'}_R}\,\psi^{j'}_L,
&\label{ffn4}\\
&\displaystyle 
{1\over f_F^2}\,\bar{\psi^j_L}\,\cD^s_R\;\bar{\cD^{s'}_R}\,\psi^{j'}_L,
&\label{ffn5}
\end{eqnarray}
which contribute to the scalar mass splittings. 

We can now discuss the scales explicitly, using naive dimensional analysis
(NDA)~\cite{nda} to estimate the effects from the topcolor scale. In NDA we
identify two scales associated with topcolor (or other generic strong
interactions). The scale $f_{TC}$ sets the scale of the composite fields
--- the amplitude to produce a composite particle is of order
$1/f_{TC}$. The scale $\Lambda_{TC}$ sets the mass scale for typical
massive strongly interacting particles (like the colorons in the topcolor
theory).\footnote{Of course, the composite scalars are lighter. Their
masses are of order $\Lambda_{TC}$ times a tunable factor that is small
because we are near the critical point where it vanishes.} Unless the
number of colors is very large, $\Lambda_{TC}$ is expected be larger than
$f_{TC}$ by a kinematic factor of about
\begin{equation}
{\Lambda_{TC}\over f_{TC}}\approx4\pi/\sqrt{N_C}\approx7\,.
\label{fourpi}
\end{equation}
All dimensional quantities in the topcolor theory can then be very roughly
estimated by assigning appropriate powers of $f_{TC}$ and
$\Lambda_{TC}$. This is by no means a calculation, but at least it gets the
kinematic factors right.

In (\ref{ffn1}) and (\ref{ffn2}), $\xi_R\;\bar{\xi_L}$ develops a vacuum
expectation value at the topcolor scale of order
\begin{equation}
\left\langle\xi_R\;\bar{\xi_L}\right\rangle
\approx\Lambda_{TC}f_{TC}^2.
\label{tcvev}
\end{equation}
Thus we expect singlet fermion masses of the order of
\begin{equation}
m_{sf}={\Lambda_{TC}f_{TC}^2
\over f_F^2}.
\label{msf}
\end{equation}

In (\ref{ffn3}), (\ref{ffn4}) and (\ref{ffn5}), the fermion bilinears get
replaced by composite scalar fields at the topcolor scale, via
\begin{equation}
\bar{\psi^j_L}\,\cU^r_R\rightarrow \Lambda_{TC}f_{TC}\,\phi.
\label{fbtos}
\end{equation}
Thus the scalar mass terms induced by interactions at the flavor scale are
of order
\begin{equation}
{\Lambda_{TC}f_{TC}\over f_F}.
\label{msflavor}
\end{equation}
Now if the theory is to produce a Higgs doublet with a negative mass
squared small compared to the common mass of the scalars, these terms must
approximately cancel the common mass. Thus we expect
\begin{equation}
m_s={\Lambda_{TC}f_{TC}\over f_F}.
\label{mscalar}
\end{equation}

We can now see how all the scales get determined. It is clear that if we
fix two of the masses, then the rest get determined. For example, in terms
of $m_{sf}$ and $f_F$, we can get $f_{TC}$ from (\ref{msf}) and
(\ref{fourpi}):
\begin{equation}
f_{TC}\approx\left({\sqrt3\,m_{sf}f_F^2\over4\pi}\right)^{1/3}.
\label{fs}
\end{equation}
Then we can compute $m_s$ and $\Lambda_{TC}$. But we can also get a bound
on $m_{sf}$ by using the renormalization group equation, (\ref{rg2}), and
the phenomenological bound from (\ref{sb}) and (\ref{cs2}). If we assume
that $m_{sf}$ saturates the bound --- that it is as small as it can be
consistent with (\ref{sb}), then all the masses are determined. What we
would hope to find is that we get $m_{sf}$ of the order of the electroweak
breaking scale, $\Lambda$. If the singlet mass are much smaller, they are
ruled out experimentally. If they are much bigger, then our analysis is not
complete. We would have to account for the difference between $m_{sf}$ and
$\Lambda$, and we would need finer tunings as well.

The result of this calculation is shown in figure~\ref{fig-scales}. The
solid lines show $f_{TC}$ and $m_{sf}$ as functions of $f_F$ assuming the
compositeness condition, (\ref{compositeness}). The dashed lines are
calculated assuming $h(\Lambda_{TC})=2$. Evidently, everywhere in this
range, $m_{sf}$ is in the appropriate range. The degree of fine-tuning
increases as $f_F$ increases, but is still less than one part in 100 (as
measured by $f_{TC}^2/f_F^2$) for $f_F=100$~TeV.

{\figsize\begin{figure}[htb]
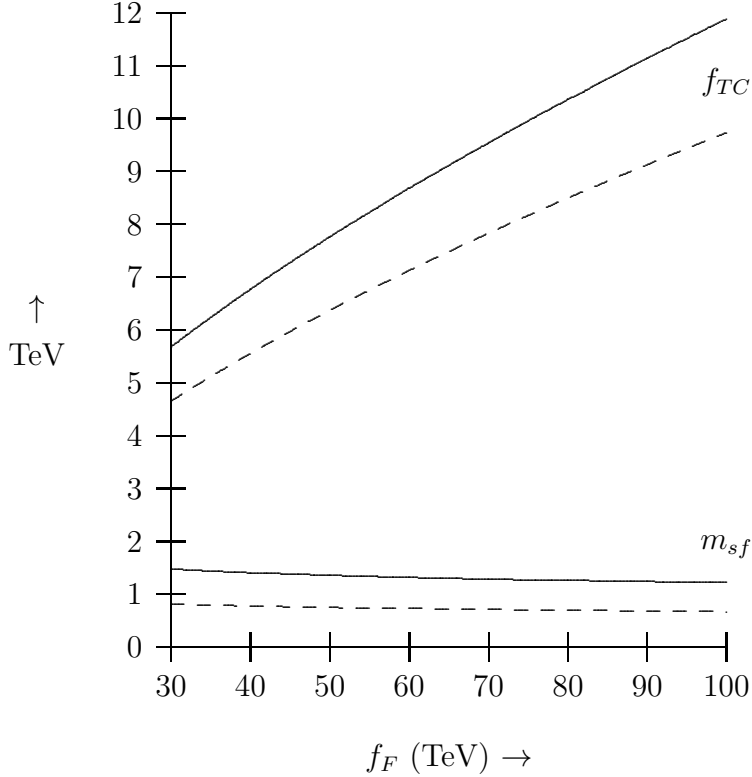

$$\beginpicture
\setcoordinatesystem units <3\tdim,20\tdim>
\setplotarea x from 30 to 100, y from 0 to 12
\axis bottom visible /
\axis left visible /
\axis bottom label {$f_F$ (TeV) $\rightarrow$} 
  ticks out short numbered from 30 to 100 by 10 /
\axis bottom ticks in short from 30 to 100 by 10 /
\axis left label {\stack{$\uparrow$,,TeV}} 
  ticks out short numbered from 0 to 12 by 1 /
\axis left ticks in short from 0 to 12 by 1 /
\visibleaxes
\normalgraphs
\plot
30 1.477 32 1.461 34 1.446 36 1.432 38 1.419 40 1.407 42 1.396 44 1.385 46
1.375 48 1.366 50 1.357 52 1.348 54 1.340 56 1.332 58 1.325 60 1.318 62
1.311 64 1.305 66 1.298 68 1.292 70 1.286 72 1.281 74 1.275 76 1.270 78
1.265 80 1.260 82 1.255 84 1.251 86 1.246 88 1.242 90 1.238 92 1.233 94
1.229 96 1.226 98 1.222 100 1.218 /
\put {$f_{TC}$} at 100 10.7
\put {$m_{sf}$} at 100 1.9
\plot
30 5.679 32 5.908 34 6.131 36 6.348 38 6.562 40 6.769 42 6.976 44 7.177 46
7.375 48 7.569 50 7.762 52 7.950 54 8.135 56 8.315 58 8.503 60 8.684 62
8.856 64 9.028 66 9.200 68 9.373 70 9.537 72 9.709 74 9.873 76 10.04 78
10.20 80 10.36 82 10.51 84 10.68 86 10.83 88 10.99 90 11.14 92 11.29 94
11.44 96 11.59 98 11.74 100 11.88 /
\setdashes
\plot
30 .8128 32 .8048 34 .7973 36 .7903 38 .7837 40 .7775 42 .7716 44 .7659 46
.7606 48 .7554 50 .7505 52 .7458 54 .7413 56 .7370 58 .7328 60 .7288 62
.7249 64 .7211 66 .7175 68 .7139 70 .7105 72 .7072 74 .7039 76 .7008 78
.6977 80 .6948 82 .6919 84 .6890 86 .6863 88 .6836 90 .6809 92 .6784 94
.6758 96 .6734 98 .6710 100 .6686 /
\plot
30 4.654 32 4.843 34 5.027 36 5.207 38 5.383 40 5.556 42 5.725 44 5.890 46
6.053 48 6.213 50 6.371 52 6.527 54 6.680 56 6.829 58 6.978 60 7.125 62
7.269 64 7.412 66 7.552 68 7.692 70 7.829 72 7.965 74 8.099 76 8.233 78
8.364 80 8.495 82 8.626 84 8.749 86 8.881 88 9.004 90 9.127 92 9.250 94
9.373 96 9.496 98 9.610 100 9.733 /
\endpicture$$
\caption{\figsize\sf\label{fig-scales} The solid lines show $f_{TC}$ and 
$m_{sf}$ as functions of $f_F$ assuming the compositeness condition,
(\ref{compositeness}). The dashed lines are calculated assuming
$h(\Lambda_{TC})=2$.}
\end{figure}}

\section{The Higgs Mass\label{higgsmass}}

In this section we study the stability of the composite Higgs potential,
and make an estimate of the composite Higgs mass.  In
\cite{Chivukula:1993pm} it was pointed out that quantum corrections to the
Higgs potential can destabilize the hierarchy between the weak scale and
the topcolor scale.  Specifically, the authors of \cite{Chivukula:1993pm}
used the renormalization group to evolve the quartic couplings of the Higgs
potential from the topcolor scale to the weak scale.  In certain cases, it
was found that the Higgs potential was unstable in the infrared.  It was
concluded that the electroweak symmetry-breaking phase transition cannot
be second order in the parameters of high-energy topcolor theory.  This
means that one cannot tune the couplings to obtain the $v\ll\Lambda_{TC}$:
as the couplings are varied, the Higgs VEV jumps discontinuously from zero
to some large value of order $\Lambda_{TC}$.  In \cite{Bardeen:1994pj}, it
was argued that in many cases the phase transition can be second order. In
this section, we will determine the range of parameters for which the phase
transition of our model is second order.

Below the topcolor scale, the theory can be described in terms of a
composite Higgs field interacting with fermions.  If we neglect the weak
$SU(2)$ and $U(1)$ interactions, the symmetry of the theory is 
\begin{equation}
SU(6)_L\times SU(6\mathord{+}n_\chi\mathord{+}n_\omega)_{U+D}
\equiv SU(N_L) \times SU(N_R),
\end{equation}
and we can write the renormalizable interactions in terms of the $N_L\times
N_R$ matrix Higgs field $H$ as (see Eq.~(\ref{yukawa}))
\begin{equation}
L_{\rm int} = h ( \bar{\psi_L} H \psi_R + {\rm h.c.} )
- \lambda_1 {\rm Tr}( H^\dagger H )^2
- \lambda_2 ( {\rm Tr} H^\dagger H )^2.
\end{equation}
The Higgs potential is bounded below if $\lambda_1$ and $\lambda_2$ obey
\begin{equation}
\label{lines}
\begin{array}{cc}
\lambda_1(\mu)+\lambda_2(\mu) > 0,~\rm{and} & 
\lambda_1(\mu) + N_L \lambda_2(\mu) > 0.
\end{array}
\end{equation}
In order for the phase transition to be second order, the couplings must
remain above these ``stability lines'' as they are evolved from the
topcolor scale down to the weak scale.

The renormalization group equations for $\lambda_1$, $\lambda_2$, and $h$
are 
\begin{eqnarray}
\frac{ d\lambda_2}{dt} &=&
\frac{1}{16\pi^2}
\left[ 4 ( N_L N_R + 4 ) \lambda_2^2 + 8 ( N_L + N_R ) \lambda_1 \lambda_2
+ 12 \lambda_1^2 + 4 N_C h^2 \lambda_2 \right],
\nonumber\\
\frac{ d\lambda_1 }{ dt } &=& \frac{1}{16\pi^2} \left[ 4 ( N_L + N_R )
\lambda_1^2 + 24 \lambda_1 \lambda_2 - 2 N_C h^4 + 4 N_C h^2 \lambda_1
\right],
\nonumber\\
\frac{ dh }{ dt } &=& \frac{1}{32\pi^2}\left[ 2 N_C + N_L + N_R \right] h^3,
\end{eqnarray}
where $t=\ln(\mu/\Lambda_{TC})$.  With the choice of scales discussed in
sec.~\ref{scales}, we have $f_F\sim 100$~TeV, $f_{TC}\sim 10$~TeV, and a
common scalar mass $m_s$ of order 10~TeV.  Below the common scalar mass, we
assume, as in sec.~\ref{fv}, that the theory contains only a single
composite doublet, and so the Higgs potential has only one quartic coupling
$\lambda$.  The coupled renormalization group equations for $\lambda$ and
$h$ are such that $\lambda$ cannot become negative as we go to lower
scales.  So if the vacuum is stable down to $\mu=m_s$, it will remain
stable down to the weak scale.  The value of $\lambda(m_s)$ depends on
details of the flavor physics: in terms of the matrices $u$ and $d$ and
appearing in (\ref{huudd}) and the angle $\theta$ appearing in
(\ref{phi2}), we have
\begin{eqnarray}
\lambda(\mu) &=& \biggl[ \lambda_2(\mu)
+ \lambda_1(\mu)\,\tr\,(\cos^2\theta u^\dagger u+\sin^2\theta d^\dagger
 d)^2 \biggr] \nonumber\\ &\leq& \biggl[\lambda_1(\mu) + \lambda_2(\mu)
 \biggr]\equiv \lambda_{+}(\mu).
\end{eqnarray}
The upper bound will suffice for our purposes, since, as we will see, the
Higgs mass turns out to be relatively light.  In the subsequent analysis,
we will use $\lambda_{+}(\mu)$ as our estimate of the Higgs quartic
self-coupling.

\begin{figure}
\centerline{\epsfxsize 4.3 truein \epsfbox{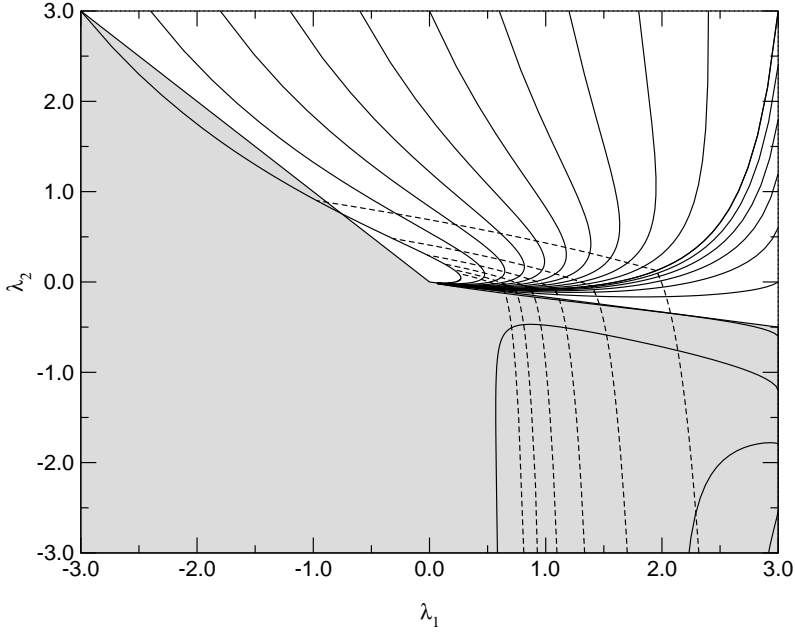}}
\caption{\figsize\sf\label{RGFigure} Renormalization group trajectories in 
the $\lambda_1$-$\lambda_2$ plane for $h(\Lambda_{TC}) = 3.5$.  The solid
curves are the trajectories, which begin at the boundary of the figure at
$\mu = \Lambda_{TC}$.  The dotted curves indicate
$\mu=\Lambda_{TC}/2,\Lambda_{TC}/4,\dots$.  In the shaded region, the
vacuum is unstable.}
\end{figure}

In fig.~\ref{RGFigure}, we display some typical renormalization group flows
in the $\lambda_1-\lambda_2$ plane.  We see that most of the trajectories
which are stable in the ultraviolet remain stable as $\mu$ is reduced from
$\Lambda_{TC}$ to $m_s$. The stability is fairly insensitive to the value
of $h$ at the topcolor scale.  If we consider all values of $\lambda_1$ and
$\lambda_2$ with $|\lambda_{1,2}(\Lambda_{TC})|<10$, we find that about 65
percent of the renormalization group trajectories are stable over the range 
$\Lambda_{TC} > \mu > m_s$ for $h(\Lambda_{TC})=2$,
90 percent are stable for $h(\Lambda_{TC})=5$, and about 30 percent are
stable for $h(\Lambda_{TC})=10$.  If we restrict the values of
$\lambda_{1,2}$ to $|\lambda_{1,2}(\Lambda_{TC})|<3$, the conclusions are
qualitatively unchanged, except that there are fewer stable trajectories
for $h(\Lambda_{TC})\gtrsim 9$.

We can use figure~\ref{RGFigure} to estimate the Higgs mass.  In terms of
$\lambda(\Lambda)$, we have
\begin{equation}
m_{\rm Higgs}^2 = 2 \lambda(\Lambda) v^2 \leq 2 \lambda_{+}(\Lambda) v^2.
\end{equation}
The surprising feature of figure~\ref{RGFigure} is that the values of
$\lambda_{1,2}$ are very strongly focussed by renormalization group flow.
As a result, we can make a qualitative estimate of the Higgs mass without
detailed knowledge of the matching conditions at the topcolor scale. If we
consider the region of parameter space $|\lambda_{1,2}(\Lambda_{TC})| < 10$
and $h(\Lambda_{TC}) \sim 2-10$, we find
\begin{equation}
\begin{array}{ccc}
-0.4<\lambda_1(m_s)<1.3,&-0.25<\lambda_2(m_s)<0.35,& 
\lambda_{+}(m_s) = \lambda_1(m_s)+\lambda_2(m_s) < 1.1.
\end{array}
\end{equation}
Evolving the rest of the way down to $\mu=\Lambda$, we find
$\lambda_{+}(\Lambda)\ltsim 0.9$, and so
\begin{equation}
m_{\rm Higgs} \ltsim 330~{\rm GeV}.
\end{equation}
This is relatively light in comparison to typical topcolor models.

\section{Conclusions\label{conclusions}}

The singlet quarks discussed in section~\ref{singlets} will decay
predominantly by GIM-violating interactions:
\begin{equation}
\chi\rightarrow cZ\,, sW^+ \quad\quad
\omega\rightarrow sZ\,, cW^-.
\label{decays}
\end{equation}
The bounds that we found in section~\ref{singlets} on the singlet masses
apply only to the singlets that mix significantly with the left-handed
doublets. At least one additional singlet is needed for anomaly
cancellation in this model, and there are no strong indirect bounds on the
masses of these non-mixing singlets. Presumably, their mixing will not be
completely absent, and they will also decay by GIM violating processes to
$W$s and $Z$s and the quarks with which they mix slightly, but with smaller
widths. This rich quark phenomenology, possibly at quite accessible
energies, is forced upon us by the constraints of anomaly cancellation in
models like those of figure~\ref{fig-3}. In addition, in contrast to many
topcolor models, we expect a relatively light Higgs. In this model, we also
expect a large family of composite doublets, but these will appear only at
the topcolor scale and may not be observable at the LHC.

\section{Acknowledgments}

We thank Markus Luty and John Terning for helpful discussions and to Sekhar Chivukula and Bogdan Dobrescu for comments on the manuscript.

\appendix
\renewcommand{\arraystretch}{1}

\section{A toy model of flavor physics\label{flavor}}

To get some experience with flavor physics of this kind, we have undertaken
a little demonstration project. We gauge the flavor symmetries in
figure~\ref{fig-4} with an $SO(3)$, and parameterize the mass terms from
(\ref{fn1}) and (\ref{fn2}) and look at the masses that result to see
whether we can get something that looks interesting. It will turn out that
we cannot find the kind of hierarchical mass matrix that we need. A more
elaborate scheme is needed, but we hope that some readers may find the
exercise useful.

We consider a model with $N_\U=3\mathord{+}n_\chi$ charge $2/3$ quarks
$\U_R^m$, $N_\D=3\mathord{+}n_\omega$ charge $-1/3$ quarks $\D_R^m$, and three
$SU(2)$ doublets of quarks $\psi_L^m$, all of which feel the topcolor
interaction. The $\U_R$ and $\D_R$ multiplets include both the standard
model up and down quarks as well as the weak singlet $\chi$ and $\omega$
quarks. The composite Higgs fields are
\begin{equation}
H_\U^{m\bar{m}} = \bar{\U}_R^{\bar{m}} \psi_L^m
\end{equation}
and
\begin{equation}
H_\D^{m\bar{m}} = \bar{\D}_R^{\bar{m}} \psi_L^m
\end{equation}
In the absence of flavor-dependent couplings, and ignoring the weak
$SU(2)\times U(1)$ interactions, the Higgs potential will have the form
\begin{eqnarray}
\label{Higgs2}
V &=& -m^2 ( \tr H_\U^\dagger H_\U + \tr H_\D^\dagger H_\D )
+ \lambda_1 \tr \left[( H_\U^\dagger H_\U )^2
+  ( H_\D^\dagger H_\D )^2 
+ 2 H_\U^\dagger H_\D H_\D^\dagger H_U \right]
\nonumber\\
&+& \lambda_2 \left[ (\tr H_\U^\dagger H_\U )^2 
+ ( \tr H_\D^\dagger H_\D )^2
+ 2\tr ( H_\U^\dagger H_\U ) \tr ( H_\D^\dagger H_\D )\right]\,.
\end{eqnarray}
This preserves the $SU(3)_L\times SU(N_\U\mathord{+}N_\D)_R$ symmetry
of the high-energy theory. 

The degeneracy of the Higgs fields can be lifted by gauging an $SO(3)$
subgroup of the flavor symmetries. As a simple example, we place the three
doublets into a triplet of $SO(3)$. We place the $\U$ and $\D$ quarks in
representations of dimension $N_\U$ and $N_\D$. The Higgs bosons then live
in product representations of $SO(3)$, which can be decomposed into
representations of different total spin.  If the $SO(3)$ is broken at a
high scale (say 100 TeV or so), then in the low energy theory, single
massive gauge boson exchange results in four-fermion operators of the form
(\ref{fn4},\ref{fn5}).  These operators lift the degeneracy of Higgs fields
with different total spin $J$.  Writing the Higgs fields in terms of states
of total $J$ using the Clebsch-Gordan decomposition,
\begin{equation}
\Phi^{JM}_\Q = \sum_{m\bar{m}}\langle JM | m \bar{m} \rangle H^{m\bar{m}}_\Q.
\end{equation}
the Higgs mass terms have the form
\begin{equation}
\label{MassTerms}
V_{\rm Mass} = \sum_{JM} ( m_1^2 + \Delta_J m_2^2 ) |\Phi^{JM}_\U|^2
 + \sum_{J'M'} ( m_1^2 + \Delta_{J'} m_2^2 ) |\Phi^{J'M'}_\D|^2,
\end{equation}
where
\begin{equation}
\Delta_J = \frac{1}{2}[ J(J+1) - j_{\Q} ( j_{\Q} + 1 ) - j_L ( j_L + 1 ) ].
\end{equation}
We expect $m_2^2>0$, so that the states of lowest $J$ have the smallest
masses and tend to condense more readily than those with higher $J$.

A further contribution to the Higgs potential comes from four-fermion
operators that mix $\chi_L$ with $\U_R$ and $\omega_L$ with $\D_R$. These
have the form
\begin{equation}
\frac{1}{M_f^2} \lambda^\U_{\bar{m}\alpha} 
(\bar{\chi}_L^\alpha \xi_R )
(\bar{\xi}_L \U_R^{\bar{m}} ) 
+
\frac{1}{M_f^2} \lambda^\U_{\bar{m}\alpha} 
(\bar{\omega}_L^\alpha \xi_R )
(\bar{\xi}_L \D_R^{\bar{m}} ),
\end{equation}
where $M_f$ is the scale of flavor physics, of order 100 TeV or so.  After
the $\xi$ fields condense, these operators yield mass terms in the low
energy:
\begin{equation}
 \mu^{\U}_{\alpha\bar{m}} \bar{\chi_L}^\alpha \U_R^{\bar{m}}
+\mu^{\D}_{\alpha\bar{m}} \bar{\omega_L}^\alpha \D_R^{\bar{m}}.
\end{equation}
These operators break the $SU(N_\U + N_\D)_R$ symmetry of the
theory. This symmetry breaking should be reflected in some way by the
Higgs potential. From the transformation properties of the symmetry
breaking parameters $\mu^\Q$, we conclude that the symmetry breaking terms
in the Higgs potential can be obtained by inserting factors of 
$(\mu^{\Q,\dagger} \mu^{\Q})_{\bar{m}\bar{m}'}$ at appropriate places in the
Higgs potential. In particular, we expect terms of the form
\begin{equation}
\label{SymmetryBreaking}
 \tr H^\dagger [\mu^{\dagger} \mu] H.
\end{equation}
These terms lift the degeneracy of Higgs fields with different ``magnetic''
quantum number $M$.

We can also fix the sign of these $H^\dagger [\mu^{\dagger} \mu] H$ terms.
In the limit that $\mu$ is large, some of the quarks become massive, and
these masses must be reflected in the spectrum of composite scalars.  As
$\mu$ gets larger, the scalars must get heavier, and we conclude that the
$H^\dagger [\mu^{\dagger} \mu] H$ contribution to the Higgs mass matrix
must be positive.  In the limit that $\mu$ is very small, this argument no
longer works.  However, we can include the effects of $\mu$ perturbatively
to reach the same conclusion.  Indeed, evaluating the correction to the
composite scalar mass from Fig.~\ref{SelfEnergy}, we find a contribution to
the Higgs mass matrix of the form
\begin{equation}
\Delta m^2 = \frac{h^2}{16 \pi^2} \mu^\dagger \mu 
\log(\Lambda_{tc}^2/\bar\mu^2),
\end{equation}
where $\bar\mu$ is a scale of order the typical singlet quark mass and
$\Lambda_{tc}$ is the scale of the topcolor interactions.  Hence in this
limit we also find that the $H^\dagger [\mu^{\dagger} \mu] H$ terms give a
positive contribution to the Higgs mass matrix.

{\figsize\begin{figure}
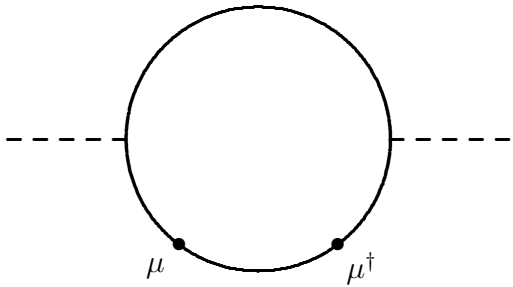

$$\beginpicture
\setcoordinatesystem units <\tdim,\tdim>
\stpltsmbl
\circulararc 360 degrees from 50 0 center at 0 0
\put {$\bullet$} at -30 -40
\put {$\bullet$} at 30 -40
\put {$\mu$} at -39 -49
\put {$\mu^\dagger$} at 39 -49
\setdashes
\plot 50 0 100 0 /
\plot -50 0 -100 0 /
\linethickness=0pt
\putrule from -100 0 to 100 0 
\putrule from 0 50 to 0 -50
\endpicture$$
\caption{\figsize\sf\label{SelfEnergy} The one-loop correction to the 
composite scalar mass with two insertions of $\mu$.}
\end{figure}}

Now we would like to determine if this model gives a seesaw mass matrix
for the quarks.  This will depend on the relative ``alignment'' of $\mu$
and $\langle H \rangle$.  The Yukawa couplings are
\begin{equation}
h \bar\psi_L H_\U \U_R +
h \bar\psi_L H_\D \D_R.
\end{equation}
After $SU(2)$ breaking, the quark mass matrix is
\begin{equation}
\left( \begin{array}{cc} \bar u^m_L \ \bar \chi^\alpha_L \end{array} \right)
\left( \begin{array}{c} h 
\langle H^{m\bar{m}} \rangle
\\ \mu_{\alpha \bar{m}} \end{array} \right)
\left( \begin{array}{c} {\U_R^{\bar{m}}} \end{array} \right).
\end{equation}
To obtain a seesaw mass matrix, we would like $\langle H \rangle$ to
align with $\mu$ such that the largest entries of $\mu$ lie in the same
column(s) as the largest entries of $\langle H \rangle$.  Unfortunately, we
find that this is not the case:  the $\mu^{\dagger}\mu$ terms lift the
masses of the Higgs fields such that the columns with the largest entries
of $\mu$ also have the most positive masses, and so do not acquire large VEVs.

To illustrate this quantitatively, take the case $n_\chi=2$, $n_\omega=1$.
For the parameters of the Higgs potential, take
\begin{equation}
\lambda_1 = 1,~\lambda_2 = 1,~m_1^2 = 1.37~{\rm TeV}^2,
~m_2^2  = 0.49~{\rm TeV}^2.
\end{equation}
For this choice of masses, only the up-type Higgses with the lowest $J$
have negative mass-squared.  For the singlet-quark mass-matrices, take (in
TeV units)
\begin{equation}
\mu^\U = \left(
\begin{array}{ccccc}
0 & 0 & 0    & 2.4 & 3.6 \\
0 & 0 & 0.49 & 3.6 & 2.4 \\
\end{array}
\right),
\end{equation}
and
\begin{equation}
\mu^\D = \left(
\begin{array}{cccc}
0 & 0 & 0 & 1 
\end{array}
\right).
\end{equation}
We include the effect of these masses on the Higgs potential by simply
adding $\mu^{\dagger} \mu$ to the Higgs mass matrix.  Minimizing the
potential, we find that $\langle H_\D \rangle = 0$, while
\begin{equation}
h \langle H_\U \rangle = 
\left(
\begin{array}{ccccc}
0     &    0    &    0.19   &    -0.05    &   0.03 \\
0     & -0.33   &    0.05   &    -0.01    &     0  \\
0.46  & -0.04   &    0.01    &      0      &     0  \\
\end{array}
\right)
\end{equation}
We have used the Pagels-Stokar estimate of the Yukawa coupling $h$ with
$\Lambda \sim 10 \mu$, resulting in $h\simeq 3.4$.  The Higgs VEV is mostly
in the $J=1$, $M=-1$ direction, with small components on the $M=0,1$
directions.  The resulting quark masses, in GeV, are
\begin{equation}
m^\U = \begin{array}{ccccc}
195, & 330, & 460, & 1230, & 5940.
\end{array}
\end{equation}
The resulting spectrum does not have light quarks that could be identified
with the first two generations, and does not yield a seesaw suppression of
the top-quark mass.  This result is generic over all of the parameter
space.

\end{document}